\documentclass[aps,prl,a4paper,twocolumn,noshowpacs,citeautoscript,10pt]{revtex4-1}

\usepackage{graphicx}

\begin{document}

\title{Observation of half-quantum vortices in topological superfluid $^3$He}
\author{S.~Autti$^{1\ast}$}
\author{V.V.~Dmitriev$^{2}$}
\author{J.T.~M\"akinen$^{1}$} 
\author{A.A.~Soldatov$^{2,3}$}
\author{G.E.~Volovik$^{1,4}$}
\author{A.N.~Yudin$^{2}$}
\author{V.V. Zavjalov$^{1}$}
\author{V.B.~Eltsov$^{1}$}

\affiliation{$^{1}$Low Temperature Laboratory, Department of Applied Physics, Aalto University, POB 15100, FI-00076 AALTO, Finland; *E-mail:   samuli.autti@aalto.fi \\
$^{2}$P.L. Kapitza Institute for Physical Problems of RAS, 119334, Moscow, Russia\\
$^{3}$Moscow Institute of Physics and Technology, 141700, Dolgoprudny, Russia\\
$^{4}$Landau Institute for Theoretical Physics, 142432, Chernogolovka, Russia.}

\begin{abstract}
One of the most sought-after objects in topological quantum matter systems is a vortex carrying half a quantum of circulation. They were originally predicted to exist in superfluid $^3$He-A, but have never been resolved there. Here we report an observation of half-quantum vortices (HQVs) in the polar phase of superfluid $^3$He. The vortices are created with rotation or by the Kibble-Zurek mechanism and identified based on their nuclear magnetic resonance signature. This discovery provides a  pathway for studies of unpaired Majorana modes bound to the HQV cores in the polar-distorted A phase.
\end{abstract}

\maketitle


Topology reveals unified features in quantum-matter systems as diverse as the chiral A-phase of superfluid $^3$He at temperatures $T \sim 10^{-3}\,$K and the vacuum of the Standard Model at $T \sim 10^{15}\,$K \cite{VolovikBook}. Those common topological properties are manifested in supported topological defects (like vortices, skyrmions etc.) and gapless quasiparticle states. Superfluid phases of $^3$He provide a versatile platform for studying topological properties of quantum matter \cite{doi:10.7566/JPSJ.85.022001,VolovikBook}. One of the most striking consequences of non-trivial topology is a vortex which carries half a quantum of circulation in a superfluid or of magnetic flux in a superconductor. Such vortices are also predicted to host core-bound unpaired Majorana modes \cite{Volovik1999,ReadGreen2000,Ivanov2001}. The search for half-quantum vortices (HQVs) began in chiral Weyl superfluid $^3$He-A in the late 70's. Despite promising theoretical predictions, HQVs in $^3$He-A remained elusive in 
experiments 
\cite{VolovikMineev1976,CrossBrinkman1977,SalomaaVolovik1985,HuMaki1987,Kawakami2009}, as the spin-orbit interaction makes them unstable \cite{Hakonen1987,Yamashita2008}. In the meantime observations of half-quantum vortices and fluxes have been reported on the grain boundaries of $d$-wave cuprate superconductors \cite{Kirtley1996}, in chiral superconductor rings \cite{HQVtriplet}, and in Bose condensates \cite{HQVpolariton,
Seo2015}. In Bose systems, however, vortex-core-bound fermionic states do not exist, while in superconductors only half-quantum fluxes not associated with a core (in Josephson vortices or rings) have so far been studied. Here we report the discovery of half-quantum vortices (HQVs) in the recently found polar phase of superfluid $^3$He \cite{PhysRevB.73.060504,PolarDmitriev}.

\begin{figure*}[tb!]
\centerline{\includegraphics[width=1 
\linewidth]{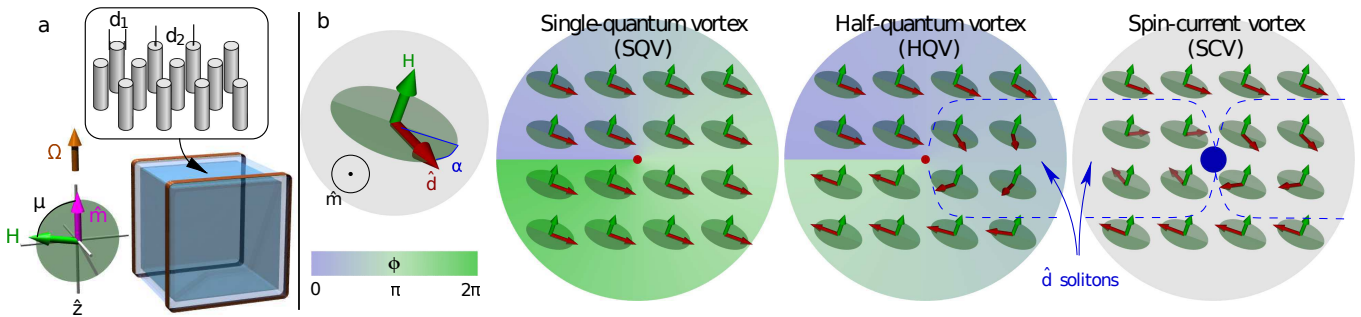}}
\caption{\label{PolarPh_vortices} ({\bf a}) Sketch of the sample: Cubic container made from epoxy (light blue) is filled with the nafen (blue cube). The nafen strands are oriented along the vertical $\hat{\mathbf{z}}$ direction, their diameter is $d_1\approx9\,$nm, and their average separation $d_2\approx35\,$nm. The space between strands is filled with liquid $^3$He. The sample is surrounded by NMR coils made of copper wire (brown rectangles). The magnetic field $ {\mathbf{H}}$ is applied transverse to the NMR coil axis at an arbitrary angle $\mu$ from the direction of the orbital anisotropy vector $\hat{\mathbf{m}}$, which is pinned along the nafen strands. The sample can be rotated around the vertical axis with the angular velocity $\Omega$ up to 3\,rad/s. ({\bf b}) Vortex types in the polar phase: The order parameter phase $\phi$ is shown by the background colour. The spin vector $\hat{\mathbf{d}}$ is locked to the plane (green disks) perpendicular to the magnetic field $\mathbf H$. Within this plane  $\
hat{\mathbf{d}}$  rotates by $\pi$ around the HQV core and by $2 \pi$ around the SCV core. In a tilted magnetic field ($\mu>0$) this winding is concentrated in one $\hat{\mathbf d}$ soliton terminating at the HQV core, or in two solitons terminating at the SCV core. The approximate soliton extents are illustrated by dashed lines. The nafen strands, the $\hat{\mathbf{m}}$ field, and the vortex lines are orthogonal to the plane of the picture. The SQV and HQV have hard cores (red discs) of the size of coherence length $\sim 40\,$nm, while the SCV has only a soft core (blue disc) of much larger size $\sim 10\,\mu$m. }
\end{figure*}

It is established that $^3$He becomes superfluid by Cooper pairing in a state with spin 1 and orbital momentum 1, which is described by a $3\times 3$ order parameter matrix $A_{\mu j}$, where indices $\mu$ and $j$ represent the spin and orbital degrees of freedom. The polar phase appears below the critical temperature $T_{\rm c}\sim 10^{-3}\,$K when liquid $^3$He is confined in nafen \cite{PolarDmitriev}, a commercially produced solid nanomaterial consisting of long strands aligned in the same direction $\hat{\mathbf{z}}$ (Fig.~\ref{PolarPh_vortices}a). In the polar phase the order parameter has the form
\begin{equation}
A_{\mu j} = \Delta e^{i\phi} \hat{d}_\mu \hat{m}_j.
\label{polar-op}
\end{equation}
Here $\Delta$ is the maximum gap in the quasiparticle energy spectrum,
$\phi$ is the superfluid phase, $\hat{\mathbf{d}}$ is a unit vector of spin anisotropy, and $\hat{\mathbf{m}}$ is that of orbital anisotropy. Magnetic field $H > 3\,$mT fixes $\hat{\mathbf{d}} = \hat{\mathbf{i}} \cos\alpha(\mathbf{r}) + \hat{\mathbf{j}} \sin \alpha(\mathbf{r})$, where $\hat{\mathbf{i}} $ and $ \hat{\mathbf{j}}$ are mutually orthogonal unit vectors in the plane normal to $\mathbf{H}$. In the nafen $\hat{\bf m}$ is pinned parallel to the strands, $\hat{\bf m} \parallel \hat{\mathbf{z}}$. The spin-orbit interaction $F_{\rm so} \propto (\hat{\mathbf{d}} \cdot \hat{\mathbf{m}})^2$ then affects orientation of $\hat{\mathbf{d}}$ in such a way that the distribution of $\alpha$ is governed by the Sine-Gordon equation \cite{SM_note}
\begin{equation}\label{SineGordonEq}
 \nabla^2 \alpha=  (\sin^2 \mu\,/\,2\xi_D^2 )\,\sin 2\alpha ,
\end{equation}
where $\xi_D\sim 10\,\mu$m is the dipole length and $\mu$ is the angle of the magnetic field with respect to $\hat{\mathbf{z}}$ (Fig.~\ref{PolarPh_vortices}a). 

A quantized vortex is a linear topological defect in the order parameter field which traps non-zero winding $\phi \rightarrow \phi+2\pi\nu$ of the order parameter phase $\phi$ on a path around the vortex core. Winding of the $\hat{\mathbf{d}}$ orientation, given by the angle $\alpha$, is also possible. Thus, three different types of vortices may exist (Fig. \ref{PolarPh_vortices}b): single-quantum vortices (SQV), spin-current vortices (SCV), and HQVs. SQVs are usual phase vortices with $\nu=1$, associated with one quantum $\kappa = h/2m_3$ of mass circulation with velocity $\mathbf{v}_{\rm s} = (\hbar/2m_3) \nabla \phi$ ($m_3$ is $^3$He atom mass). In a SCV $\alpha \rightarrow \alpha+2\pi$ on a path around the core. It carries quantized circulation of spin current, created by reorientation of the $\hat{\mathbf d}$ field. For a HQV $\nu=1/2$, and it carries only half a quantum $\kappa/2$ of mass circulation. The apparent phase jump by $\pi$ is compensated by the change $\alpha \rightarrow \alpha+\pi$, which 
makes the order parameter single valued.

Reorientation of $\hat{\mathbf{d}}$ outside of SCV and HQV cores is governed by well-known solitonic solutions of Eq.~(\ref{SineGordonEq}). The SCV becomes a termination line of two and HQV of one $\pi$ soliton in the $\hat{\mathbf d}$ field. The soliton connects a HQV to another HQV with opposite $\hat{\mathbf d}$ winding (SM, Fig.~S1), to a SCV or terminates at the sample boundary. The soliton width is $\sim \xi_D / \sin \mu$. In the absence of magnetic field or when the field is oriented along the nafen strands ($\mu=0$), the spin-orbit energy $F_{\rm so}$ is invariantly at the minimum for any  $\hat{\mathbf d} \perp \hat{\mathbf m}$, and solitons are not created.

In our experiments we use continuous-wave nuclear magnetic resonance (NMR). In the superfluid state the spin-orbit coupling provides a torque acting on the precessing magnetisation, which leads to a shift of the precession frequency from the Larmor value  $\omega_{\rm L}=|\gamma| H$, where $\gamma=-2.04\cdot10^8\,$s$^{-1}$T$^{-1}$ is the gyromagnetic ratio of $^3$He. The sample regions where spin-orbit energy is at a minimum form the main peak in the NMR spectrum at the frequency $\omega_{\rm main}$ with the shift \cite{Mineev2016}
\begin{equation}
\Delta \omega_{\rm main}= \omega_{\rm main}  - \omega_{\rm L}  = 
\frac{\Omega_P^2}{2\omega_L}  \cos^2\mu  .
\label{SatelliteGeneral1}
\end{equation}
Here $\Omega_P$ is the pressure- and temperature-dependent Leggett frequency in the polar phase, which characterises the spin-orbit torque.

Within soft cores of topological objects, such as the $\hat{\mathbf{d}}$ solitons, the spin-orbit energy is not at minimum, which provides a trapping potential for standing spin waves. Excitation of these waves leads to a satellite peak in the NMR spectrum at the frequency $\omega_{\rm sat}$ with the shift
\begin{equation}
\Delta \omega_{\rm sat}= \omega_{\rm sat}  - \omega_{\rm L}  =  
\frac{\Omega_P^2}{2\omega_L} (\cos^2 \mu - \lambda \sin^2\mu )  \,.
\label{SatelliteGeneral2}
\end{equation}
Here $\lambda(\mu)$ is specific to the type of the topological object. For an infinite planar $\hat{\mathbf{d}}$ soliton one has $\lambda =1$, corresponding to the zero mode on the soliton~\cite{vollhardt2013superfluid}. In this case one finds $\Delta \omega_{\rm sat} (\mu=\pi/2) = -\Delta \omega_{\rm main}(\mu=0)$. For a $\hat{\mathbf{d}}$ soliton, separating two HQVs, the signal is modified due to finite-size effects and also due to the Aharonov-Bohm effect for spin waves scattered on HQV cores (see SM). In this respect, the HQV can be seen as an analogue of the Alice string in particle physics \cite{ShouChengZhang2010}. A particle that moves around an Alice string flips its charge~\cite{Schwarz1982}, while a quasiparticle going around a HQV flips its spin quantum number. As a result, we expect $\lambda < 1$. The spin polarisation of the HQV core predicted in Ref.~\citenum{PhysRevLett.103.057003} does not affect the signal, since for two HQVs bound by the soliton, the winding of $\hat{\mathbf{d}}$ (and 
thus 
spin polarisation) is in the opposite directions.

\begin{figure}[t]
\centerline{\includegraphics[width=1\linewidth]{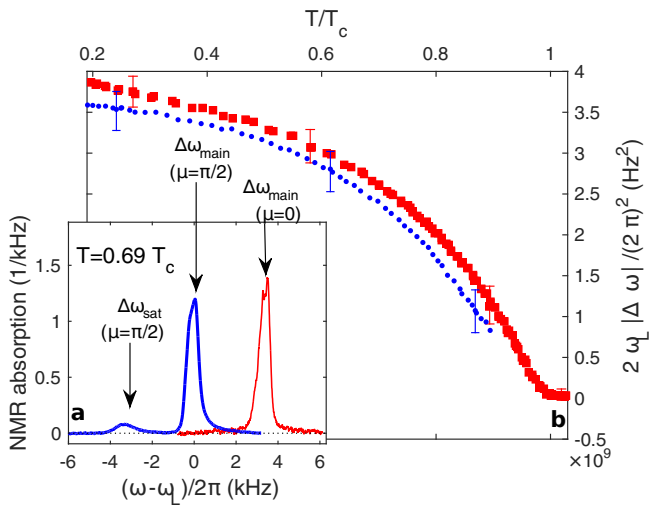}}
\caption{\label{SatelliteT}   NMR spectra in the polar phase with HQVs. (\textbf{a}) Normalised spectra measured in transverse field $\mu=\pi/2$ (blue thick line) shows the HQV satellite at the negative frequency shift $\Delta\omega_{\rm sat}$ and the main line at zero frequency shift. In the axial field $\mu=0$ (red thin line) only the main line at positive shift $\Delta\omega_{\rm main}$ is seen. This spectrum is not sensitive to presence of vortices. (\textbf{b}) Temperature dependencies of the satellite position in the transverse field $|\Delta\omega_{\rm sat}(\mu=\pi/2)|$ (blue circles) and the main line position in the axial field $\Delta\omega_{\rm main}(\mu=0)$ (red squares) closely match as expected for HQVs. The example error bars show full width at half maximum of non-Lorentzian main line as an estimate of possible systematic error. }
\end{figure}

In a sample, rotating with angular velocity $\Omega$, the lowest-energy state is achieved when superfluid mimics solid-body rotation with an array of rectilinear vortex lines, oriented along the rotation direction, with density $n_{\rm v} = 2\Omega/(\nu\kappa)$. In the polar phase the spin-orbit interaction favours HQVs \cite{Mineev2014}, and they are energetically preferable below $T_{\rm c}$ in the axial field (see SM). However, when solitons are formed between HQV pairs in the tilted magnetic field, excess of the spin-orbit interaction energy within solitons makes SQVs preferable. Presumably SCVs can be created during cooldown when strong time-dependent magnetic field is applied to generate a random distribution of vector $\hat{\mathbf d}$ \cite{Dmitriev2010}. Such experiments are beyond the scope of this work. 

We approach the lowest-energy state by slowly cooling the sample in rotation from above $T_{\rm c}$ to the polar phase. The axis of rotation is parallel to the nafen strands (see Fig.~\ref{PolarPh_vortices}a). Results presented in this work were obtained using 94\% open nafen (density 243~mg/cm$^3$), at 7.1~bar pressure and magnetic field 12~mT. The corresponding NMR frequency is $\omega_L /2 \pi=$~374~kHz. Temperature is measured from the NMR spectrum of bulk B phase, using the known B-phase Leggett frequency \cite{Thuneberg2001,Hakonen1989}. As a secondary thermometer we use a quartz tuning fork~\cite{2007_forks, 2008_forks} located in bulk ${}^3$He. Details of the experimental setup and methods are in the supplementary material \cite{SM_note}.

In the experiment when the magnetic field is oriented transverse to the nafen strands during cooldown, we observe only the bulk NMR line of the polar phase, which is consistent with the creation of SQVs. SQVs do not produce a potential well for spin waves and thus yield no satellite. When we apply no magnetic field or, alternatively, orient the field along the nafen strands during cooldown, and then turn the field to the transverse direction below $T_{\rm c}$, a satellite in the NMR spectrum is observed, Fig.~\ref{SatelliteT}a. We interpret this satellite as produced by $\hat{\mathbf{d}}$ solitons connecting pairs of HQVs based on the following arguments.

\begin{figure}[tb!]
\includegraphics[width=1 \linewidth]{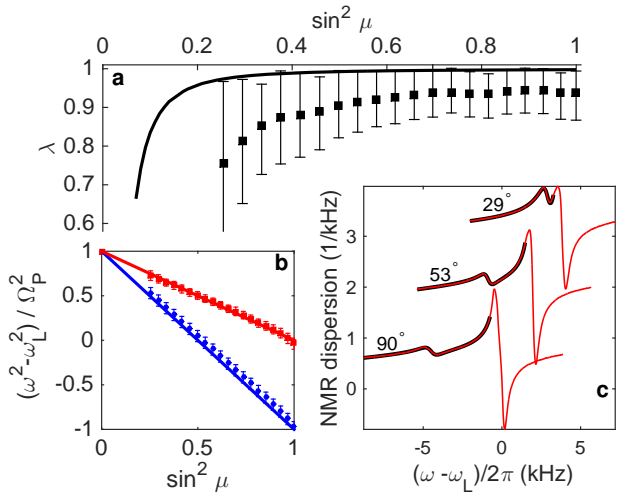}
\caption{\label{SatelliteTilt}   Frequency shift of the HQV satellite in the NMR spectrum. (\textbf{a}) Measured values of the dimensionless frequency shift $\lambda$ as a function of the field tilt angle $\mu$ (symbols), and numerical calculations for the uniform polar phase (solid line) using theoretical value of $\xi_D=17 \mu$m  \cite{SM_note}. Leggett frequency $\Omega_P$ is determined from a separate measurement at $\mu=0$.  As expected, the deviation from the infinitely-long $\hat{\mathbf{d}}$ soliton value $\lambda = 1$ increases towards small $\mu$. The disagreement between the experiment and calculations probably originates from disorder in the nafen strand orientation \cite{Asadchikov2015}, which leads to fluctuations of spin-orbit interaction energy within the solitons. (\textbf{b}) Values of $\lambda$ are found from positions of the HQV satellite $\Delta\omega_{\rm sat}$ (blue circles) and of the main line $\Delta\omega_{\rm main}$ (red squares). The red and blue solid lines 
show 
results of Eqs. (3) and (4), respectively, for $\lambda = 1$. (\textbf{c}) Main line and satellite line positions are 
determined using Lorentzian fits (thick black line) of the dispersion signal (red line). For clarity, the lines have been shifted vertically. Values of $\mu$ are marked above the lines. Similar fits are used in extracting the satellite line intensity in  Fig.~\ref{intensityOmega_KZ}. In panels (\textbf{a}) and (\textbf{b}) the bars show full width at half maximum of the spectral lines.}
\end{figure}

First, the frequency of the satellite line conforms to expectations for a $\hat{\mathbf{d}}$ soliton. In Fig.~\ref{SatelliteT} the frequency shift $-\Delta\omega_{\rm sat} (\mu=\pi/2)$ is compared to $\Delta \omega_{\rm main}(\mu=0)$ as a function of temperature. They are seen to closely match each other and we find $\lambda(\mu=\pi/2) = 0.93\pm 0.07 $. Deviation from ideal soliton value $\lambda=1$ is expected to increase with decreasing field angle $\mu$ due to growing soliton width. This is qualitatively seen in Fig.~\ref{SatelliteTilt}. 

In the analysis we assume that the distance between vortex pairs, defining the soliton length, is independent of the angle $\mu$, i.e.,\ that the tension of the soliton is not able to overcome the pinning of HQV hard cores of the size of coherence length $\xi \sim 40\,$nm on the nafen strands of $\sim 10\,$nm diameter. The strong pinning is confirmed by the observation that after stopping the rotation the satellite in the NMR spectrum remains unchanged for days, while the Magnus force, pulling vortices towards sample boundary for annihilation, exceeds the soliton tension by a factor of $10^3$.

\begin{figure}[tb!]
\centerline{\includegraphics[width=1 \linewidth]{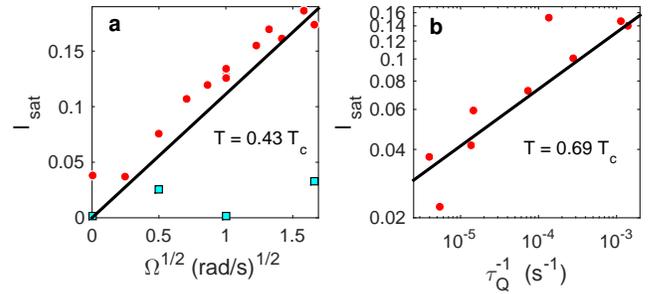}}
\caption{\label{intensityOmega_KZ} Intensity $I_{\rm sat}$ of the HQV satellite in the NMR spectrum. (\textbf{a}) After cooldown in zero field (red circles)  $I_{\rm sat}$ follows expected dependence for HQVs $I_{\rm sat}\propto \Omega^{1/2}$. The solid line is theoretical prediction for the equilibrium array of HQVs ignoring Kibble-Zurek vortices and with no fitted parameters. For cooldowns in the transverse field (cyan squares) formation of HQVs is suppressed. Residual intensity is attributed to the possible formation of the soliton sheets terminating at the sample walls. The cooldown rate here is $\tau_Q^{-1}=5\cdot 10^{-6}\,$s$^{-1}$. (\textbf{b}) For cooldowns at $\Omega=0$~rad/s the satellite intensity (symbols) follows dependence $I_{\rm sat} \propto \tau_Q^{-1/4}$ (solid line) expected for vortices created by the Kibble-Zurek mechanism, where the initial vortex density agrees with earlier measurements in $^3$He-B \cite{Bauerle1998}. In all cases spectra are measured at $\mu = \
pi/2$, and $I_{\rm sat}$ is determined as the 
area of the satellite normalised to the total area of the spectrum. }
\end{figure}

The second observation in support of HQV is the dependence of the relative satellite peak intensity $I_{\rm sat}$ on the angular velocity $\Omega$ of rotation, which is applied during the cooldown. The measured satellite intensity at $\mu=\pi/2$ is plotted in Fig. \ref{intensityOmega_KZ}a. For solitons of the $\xi_D$ width between pairs of vortices, the expected signal is  $I_{\rm sat} = (n_{\rm v}/2)\, g_{\rm s} L \xi_D$ \cite{HuMaki1987}. Here $L = b n_{\rm v}^{-1/2}$ is the average soliton length, where $g_{\rm s}\sim 1$ is a numerical factor, which depends on the distribution of the trapped spin wave within the soliton. Numerical factor $b\sim1$ depends on the vortex lattice. For very low vortex density and long solitons $L \to \infty$ one has $g_{\rm s} \to 2$. As a result, one expects $I_{\rm sat} \propto \Omega^{1/2}$, as indeed seen in Fig.~\ref{intensityOmega_KZ}a. Comparison of the measurements with the theoretical prediction for $I_{\rm sat}$ \cite{SM_note} shows good quantitative agreement, 
considering absence of fitting parameters and simplicity of the model, which assumes uniform polar phase and uniform distribution of HQVs.

A remarkable feature seen in Fig.~\ref{intensityOmega_KZ}a is that the satellite appears in the zero-field cooldowns in the absence of rotation as well. We attribute this phenomenon to the Kibble-Zurek mechanism (KZ) of vortex (defect) formation \cite{Kibble1976,Zurek1985} during the crossing of the 2nd order phase transition to the polar phase. The KZ mechanism is expected to create various order-parameter defects including vortices of all possible types. In the earlier observations of the KZ mechanism in superfluid $^3$He-B initially formed vortices rapidly decay \cite{KZ_nature,KZ_nature2,PhysRevB.90.024508,ProgLowTempPhys_page9}. In our case the initially formed HQVs freeze due to the strong pinning. The scale for the inter-vortex distance is set by the KZ length $l_{\rm KZ} = \xi_0 (\tau_Q/\tau_0)^{1/4}$. Here   $\tau_Q^{-1}=\left. -\frac{d(T/T_{\rm c})}{d t} \right|_{T=T_{\rm c}}$ is the cooldown rate at $T_{\rm c}$, $\xi_0 = \xi(T=0)$ and the order-parameter relaxation time $\tau_0 \sim 1\,$ns. For 
HQVs the inter-vortex distance sets the length of the interconnecting solitons and thus the amplitude of the satellite signal. The resulting dependence $I_{\rm sat}\propto l_{\rm KZ} \propto \tau_Q^{-1/4}$ is indeed observed in the experiment, Fig.~\ref{intensityOmega_KZ}b. The magnitude of the signal corresponds to the averaged soliton length of $1.4~ l_{\rm KZ}$, as has been estimated as the initial inter-vortex distance also in the B phase of ${}^3$He \cite{Bauerle1998,ProgLowTempPhys_page9}. The shift of experimental data in Fig.~\ref{intensityOmega_KZ}a above the theoretical expectation indicates that the KZ mechanism is important also in cooldowns with applied rotation. Detailed analysis of the creation of various defects by the KZ mechanism in the presence of bias (rotation) and pinning remains a task for the future.

In conclusion, we cool ${}^3$He within nematically ordered aerogel-like material called nafen down to the superfluid polar phase in rotation up to 2.75 rad/s. When the cooldown proceeds without magnetic field or with the field oriented parallel to the nafen strands and rotation axis, and afterwards the field is switched on in the transverse direction, we observe a satellite peak in the continuous-wave NMR spectrum. Dependence of the satellite intensity and frequency on the rotation velocity, temperature, and field orientation identifies it as a signal from solitons between pairs of half-quantum vortices. Additionally, the frequency of the satellite reflects the Aharonov-Bohm effect for the spin waves scattered by the HQV core, which is the analogue of the transfer of the Cheshire charge in Alice electrodynamics \cite{ShouChengZhang2010}. The hard-core HQVs are strongly pinned by the nafen strands, which allows us to observe vortices created when passing through the transition temperature owing to the Kibble-
Zurek 
mechanism even without rotation. If cooldown proceeds with the field applied in the transverse direction favouring stabilisation of single-quantum vortices, no satellite is observed as expected.

In less dense nafen (98\% open, 90~mg/cm$^3$) the polar phase acquires axial distortion at low temperatures, i.e. it is transformed into the polar distorted A-phase \cite{PolarDmitriev}. The HQVs there contain unpaired Majorana modes in their cores similar to those discussed in HQVs in $p_x+ip_y$ superconductors or in ``Kitaev chains'' \cite{Volovik1999,ReadGreen2000,Ivanov2001}. Since the HQVs are trapped by the nafen columnar defects, the gap separating the Majorana mode from the other Caroli-de Gennes-Matricon excitations localised in the vortex core increases from the usual minigap value $\sim \Delta^2/E_F \sim 10^{-3}\Delta$  (here $E_F$ is the Fermi energy) in an unpinned vortex to a significant fraction of the gap amplitude $\Delta$ for a vortex trapped on a defect~\cite{Melnikov2009,Shapiro2013}. Our preliminary results of HQV measurements in the 98\% open nafen show that HQVs survive the transition from the polar phase to the polar-distorted A-phase. This opens a pathway for probing the core-bound 
unpaired Majorana modes in the experimentally accessible temperature range in the polar-distorted A-phase,  and also in the new topological phases obtained using properly engineered nanostructured confinement \cite{SM_note}, while the Kibble-Zurek mechanism makes the HQVs accessible with less specialised, stationary equipment.

This work has been supported by the Academy of Finland (project numbers 
284594 and 298451), and RFBR grant 16-02-00349. We used facilities of the Low Temperature Laboratory infrastructure of Aalto 
University. S.A. acknowledges financial support from the Finnish 
Cultural Foundation. We thank I.M.Grodnensky for providing the nafen samples. We dedicate this work to an invaluable colleague and collaborator, Tom Kibble, who has recently passed away.

\renewcommand{\thefigure}{S\arabic{figure}} 
\setcounter{figure}{0}
\renewcommand{\theequation}{S\arabic{equation}} 
\setcounter{equation}{0}


\onecolumngrid
\newpage
\section*{Supplementary Material}
\twocolumngrid

\begin{figure*}[tb!]
\centerline{\includegraphics[width=0.95 \linewidth]{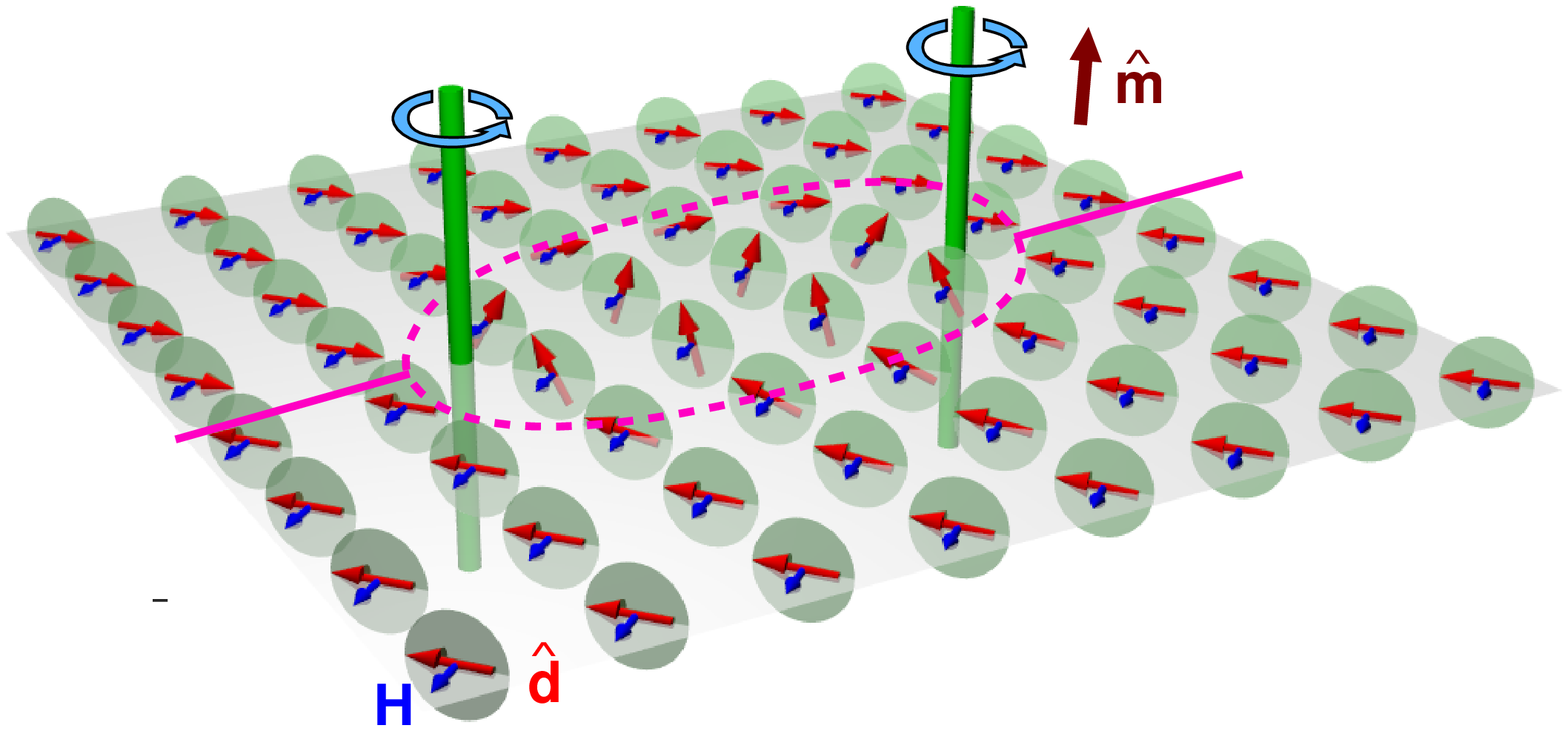}}
\caption{\label{image:HQV_soliton} A pair of HQVs (green pillars) in the polar phase of superfluid $^3$He in transverse magnetic field (blue arrows), connected by a soliton in the $\hat{\mathbf{d}}$ field (red arrows). The $\pi$ winding of $\hat{\mathbf{d}}$ around a HQV core is in the opposite direction for the two vortices. The soliton width (highlighted by the dash line) is determined by the spin-orbit interaction and is of the order of $\xi_D \sim 10~\mu$m, while the intervortex separation $\sim 100~\mu$m is set by the rotation velocity applied during the transition through $T_c$. Solid lines indicate branch cuts where $\hat{\mathbf{d}} \rightarrow -\hat{\mathbf{d}}$ and the superfluid phase jumps by $\pi$, but the order parameter is continuous. }
\end{figure*}

\section*{Experimental Setup and Methods}

The $^3$He sample is confined in a $4\times4\times4~\rm{mm}^3$ cubic container (see Figure \ref{PolarPh_vortices}a) made from Stycast 1266 epoxy. The container volume is filled with aerogel-like material called nafen of 243\,mg/cm$^3$ density. The nafen consists of strands made from Al$_2$O$_3$, oriented in the same direction. The average diameter of the strands is measured with X-ray scattering as $d_1 = 9\,$nm and variations of the diameter are about $\pm2$\,nm \cite{Asadchikov2015}. The average distance between strands, determined from the nafen density, is $d_2 \approx 35\,$nm. The nafen was produced by AFN technology Ltd in Estonia. The direction of the strands is vertical and further denoted as $\hat{\mathbf{z}}$. The sample is cooled by the ROTA nuclear demagnetisation refrigerator~\cite{Heikkinen2014}, which can be rotated around its vertical axis.

Results presented in this work were obtained at 7.1~bar pressure and magnetic field 12~mT. The corresponding NMR frequency is $\omega_L /2 \pi=$~374~kHz. The container walls and nafen strands have been preplated with a few atomic layers of ${}^4$He in order to suppress the paramagnetic signal from solid ${}^3$He otherwise formed on all surfaces.  At this pressure we have measured the transition temperature to polar phase $T_{\rm c} = (0.9 \pm 0.03) T_{\rm c}^{\rm{bulk}}=$ 1.55~mK, which is slightly lower than that reported in Ref.~\citenum{PolarDmitriev} for the same nominal nafen density. The difference is probably mostly due to variations of the actual density of the nafen.

The sample is probed using continuous-wave nuclear magnetic resonance spectroscopy (cw-NMR). The excitation and signal pick-up is performed with the coils made from copper wire which are part of a tuned tank circuit with a $Q$ value of 150. To improve signal to noise ratio a cold preamplifier thermalised to helium bath is used. The NMR excitation is applied at the fixed resonance frequency of the tank circuit and the magnitude of the magnetic field is swept to record NMR spectra. The spectra are averaged over several tens of sweeps in order to reduce the noise, except for temperature sweeps where averaging is not used. The satellite line position and amplitude, and the main line position are found by fitting Lorentzian lines to the dispersion signal (see Fig. \ref{SatelliteTilt}c), which contains less noise than the absorption.

Temperature is measured from the NMR spectrum of bulk B phase, using known B-phase Leggett frequency. The bulk response is picked up by the NMR coils from the gaps around the nafen sample and from the filling line. It is clearly visible at temperatures above 0.85$T_{\rm c}$. At lower temperatures the bulk line becomes so wide that the signal disappears in the noise, and we use as a thermometer the position of the NMR line in the polar phase in the axial field. We convert the position to the temperature using the model suggested in Ref.~\citenum{PolarDmitriev}, Eq. (4), where parameters $K=1.2$ and $T_{\rm c}=0.9~T_{\rm c}^{\rm{bulk}}$ are found by comparison to the bulk B-phase temperature at $T>0.85~T_{\rm c}$. 
As a secondary thermometer we use a quartz tuning fork~\cite{2007_forks, 2008_forks} located in bulk 
${}^3$He in a volume connected by a sintered heat exchanger to the nuclear demagnetisation refrigerator. At higher temperatures the fork is calibrated from NMR. At the lowest temperatures one can also use the onset of the ballistic regime as an approximate reference point \cite{Todo_drydemag2014}. The fork we use has the cross-section of a tine $0.4 \times 0.3 ~\rm{mm}^2$ and the onset takes place at approximately 0.28$T_{\rm c}$~\cite{JLTP48_Ono_etal}. This yields for the lowest temperature reached in the experiments $(0.19 \pm 0.05)\,T_{\rm c}$, while the lowest temperature according to the main NMR line position is $(0.2\pm 0.04)\,T_{\rm c}$.

The magnetic field can be freely rotated in the plane perpendicular to the NMR coil axes. This includes field orientations parallel to the strands and the rotation axis ($\mu=0$), and transverse to them ($\mu=\pi/2$). The direction $\mu=\pi/2$ corresponds to the maximum frequency shift of the satellite $\Delta\omega_{\rm sat}$, see Eq.~(\ref{SatelliteGeneral2}). From this condition we have found that the container axis $\hat{\mathbf{z}}$ is tilted by $1^\circ$ from the axis of the magnet assembly. Comparison of the NMR spectra at the orientations $\mu$ and $\mu + \pi$ allows us to find the magnitude of the magnetic field trapped in the superconducting parts of the setup. Those have turned out to be $40\,\mu$T along $\hat{\mathbf{z}}$, and about $2\,\mu$T in the transverse direction. The container tilt and the trapped field are taken into account in the conversion of the NMR spectra to the frequency domain.

\section*{Energy of Quantized Vortices in the Polar Phase}

In a rotating sample the lowest-energy state is achieved when superfluid mimics solid-body rotation with an array of rectilinear vortex lines. Due to quantized nature of vorticity, however, this simulation of solid-body rotation cannot be complete. The main contribution to the excess energy above the energy of the rotating solid body comes from the superflow $v_{\rm s}$ around the vortex core which diverges as $v_{\rm s} = \nu\kappa/(2\pi r)$, where $r$ is the distance from the vortex axis. If there are spin currents around the core, like in the case of HQV, then they also contribute to the excess energy. 

The divergence of the superflow is cut at the size of the vortex core. In the polar phase of superfluid $^3$He both SQV (with $\nu=1$) and HQV (with $\nu=1/2$) have hard core of the temperature dependent coherence length size $\xi(T)$. Thus their energies per circulation quantum in the logarithmic approximation become \cite{SalomaaVolovik1987}

\begin{equation}
 E_{\rm SQV} =  \frac{\kappa^2}{4\pi}\, \rho_{\rm s}^\perp \ln \frac{L}{\xi}
\end{equation}
and
\begin{equation}
  E_{\rm HQV} \big/\frac{1}{2} = \frac{\kappa^2}{4\pi}\,  \frac{1}{2} (\rho_{\rm s}^\perp + \rho_{\rm sp}^\perp) \ln \frac{L}{\xi},
\end{equation}
where $\rho_{\rm s}^\perp$ is the transverse component of the superfluid density and $\rho_{\rm sp}^\perp$ is the transverse spin rigidity (following the notation of Ref.~ \citenum{SalomaaVolovik1987}).

The Fermi-liquid effects in superfluid $^3$He make $\rho_{\rm sp}^\perp < \rho_{\rm s}^\perp$, see p. 198 in Ref.~\citenum{vollhardt2013superfluid}, with a typical value $\rho_{\rm sp}^\perp \approx 0.5 \rho_{\rm s}^\perp$ at $T = 0.5 T_{\rm c}$. Thus in the absence of $\hat\mathbf{d}$ solitons, an array of HQVs is more energetically favourable than array of SQVs, $2 E_{\rm HQV} < E_{\rm SQV}$. The ratio $\rho_{\rm sp}^\perp / \rho_{\rm s}^\perp$ grows with increasing temperature and the situation becomes less clear when $T \to T_{\rm c}$ and $\rho_{\rm sp}^\perp \to \rho_{\rm s}^\perp$. Formally the energies of HQV and SQV arrays in the logarithmic approximation become equal, but the approximation itself ceases to be valid, since $\xi$ diverges close to $T_{\rm c}$. To determine accurately the energy of vortices one then has to resort to numerical calculations of the order-parameter distribution in the core.

Moreover, since the cooldown in the experiment proceeds with a finite rate, the vortex formation is not completely equilibrium process. In this case, to calculate the proportion of vortices of each type to be formed, one should also take into account the probability to form a specific order-parameter configuration in a non-equilibrium transition through $T_{\rm c}$. These probabilities in general are not directly related to the equilibrium energies, as was discussed in the context of the transition from the normal to the A or B phase of superfluid $^3$He \cite{Bunkov_transition} and can be only estimated in numerical simulations, which remain a task for a future research. The clear experimental result, though, is that most of the vortices formed in the transition to polar phase in rotation in zero magnetic field are half-quantum.

When solitons are formed between HQV pairs in the tilted magnetic field, the excess of the spin-orbit interaction energy within solitons makes configuration with HQVs to be energetically unfavourable at all temperatures. This conclusion is also confirmed by the experiment.

\section*{NMR in Polar Phase of $^3$He}

The NMR properties of the polar phase, determined by the dynamics of 
the vector $\hat{\bf d}$, are similar to that of $^3$He-A \cite{Mineev2014}. There are three components of the Hamiltonian which are important for spin dynamics: magnetic energy,  spin-orbit interaction energy, and
gradient energy:
\begin{equation}\label{eq:ham}
\mathcal{H} = F_M + F_{SO} + F_\nabla,
\end{equation}
\begin{eqnarray}
\label{eq:en_m0}
F_M &=& - ({\bf S} \cdot \gamma {\bf H})
+ \frac{\gamma^2}{2}\chi_{ab}^{-1} S_a S_b,\\
\label{eq:en_d0}
F_{SO}
&=& g_D \Big[
           A^*_{jj}A_{kk}
         + A^*_{jk}A_{kj}
 - \frac23 A^*_{jk}A_{jk}\Big],\\
\label{eq:en_g0}
F_\nabla
&=&  \Big[
    K_1 (\nabla_j A^*_{ak})(\nabla_j A_{ak})\\\nonumber
&+& K_2 (\nabla_j A^*_{ak})(\nabla_k A_{aj})
+   K_3 (\nabla_j A^*_{aj})(\nabla_k A_{ak}) \Big],
\end{eqnarray}
where $\bf S$ is spin. Susceptibility~$\chi$ is anisotropic, the axis of anisotropy is~$\bf \hat d$,
and therefore minimum of the magnetic energy corresponds to $\bf S \perp {\hat d}$.

Let us for simplicity consider the case uniform in the ${\bf\hat z}$ direction. We can parametrise the magnetic and order parameter fields:

\begin{eqnarray}\label{eq:angles}
&&{\bf \hat m} = {\bf\hat z} 
,\\\nonumber
&&{\bf H} = {\bf\hat h} H, \qquad {\bf\hat h}= ({\bf\hat y} \sin\mu + {\bf\hat z} \cos\mu)
,\\\nonumber
&&{\bf \hat d} = ({\bf\hat i} \cos\alpha
 + {\bf\hat j} \sin\alpha)\sin\beta +{\bf\hat h}\cos\beta.
\end{eqnarray}
Here $\beta$ is angle between $\bf\hat d$ and the field, and $\alpha$ is the azimuthal angle of $\bf \hat d$ in the plane perpendicular to the magnetic field. As before, we choose $\alpha=0$ along the line perpendicular to both~$\bf H$ and~$\bf \hat m$ which corresponds to the minimum of energy. That is, ${\bf\hat i}={\bf\hat x}$ and ${\bf \hat j} = {\bf \hat y} \cos \mu - {\bf \hat z} \sin \mu $. Omitting constant terms, the energies now become

\begin{eqnarray}
\label{eq:en_m2}
F_M &=&
\frac12(\chi_\perp - \chi_\parallel)H^2 \ \cos^2\beta,
\\
\label{eq:en_d2}
F_{SO}
&=& 2g_D\Delta^2
\ (\sin\alpha\sin\beta \sin\mu + \cos\beta \cos\mu)^2,
\\
\label{eq:en_g2}
F_\nabla
&=& \Delta^2 K_{jk}
\ [\sin^2\beta (\nabla_j\alpha)(\nabla_k\alpha) \\\nonumber
&+& (\nabla_j\beta)(\nabla_k\beta)],
\end{eqnarray}
where $K_{jk} = K_1 \delta_{jk}$.

There are two scales introduced by these energies. The ratio of the magnetic and
gradient energies gives the magnetic healing length~$\xi_H$ and the ratio of the
spin-orbit and gradient energies gives the dipolar length~$\xi_D$:
\begin{equation}
\xi^2_{H} = \frac{2 K_1\Delta^2}{H^2 (\chi_\perp - \chi_\parallel)}
,\quad
\xi^2_{D} = \frac{K_{1}}{2 g_D}.
\end{equation}

We use bulk values from Ref.~\citenum{thuneberg_texture} to estimate $\xi_D$: $K_1$ scales according to $K_1 \propto 1/T_{\rm c}$ wrt. its bulk value (see Ref. \citenum{Thuneberg_aerogel}), and ignoring the additional scattering due to the nafen we get $\xi_D = 17~\mu$m. Note that our definition of $\xi_D$ follows the notation of Ref.~\cite{SalomaaVolovik1987}, which is smaller by $\sqrt{2}$ than that in e.g. Ref.~\cite{thuneberg_texture}. 

In the high-field limit $\xi_D\gg\xi_H$, that is, magnetic energy is in the minimum everywhere excluding small regions of the~$\xi_H$ size (for example, cores of spin vortices). The small volume of this regions makes
them invisible in NMR experiments. In the rest of the
volume~$\beta=\pi/2$, and only variations of~$\alpha$ are important. Minimising the total energy wrt. $\alpha$ yields Eq.~(\ref{SineGordonEq}). In a tilted magnetic field there are two possible uniform equilibrium textures with~$\alpha=0$ and~$\alpha=\pi$, that is, vector~$\bf d$ is perpendicular to both~$\bf H$ and~$\bf l$ and
can point in two possible directions. The two states may coexist, and in between there is a quasi-stable $\bf d$ soliton. The thickness of such soliton can be solved analytically and is of the order of $\xi_D$. The soliton may terminate e.g. at a container wall, at a half-quantum vortex, or at a spin vortex.

\section*{HQVs and NMR}

The NMR response of bulk polar phase can be derived by considering tiny oscillations of the spin part of the order parameter field. The position of the satellite peak, derived in a similar fashion, is given by the eigenvalue $\lambda$ of the wave equation describing the spin wave modes localised in the texture of the vector $\hat{\bf d}$ (see Fig.~\ref{image:HQV_soliton}). Note that we define $\lambda$ so that it is positive for practical reasons. Assuming vortices are oriented along $\hat{\bf{ z}}$, it acquires 
the following form \cite{SalomaaVolovik1987}:
\begin{eqnarray}
-\lambda \Psi = \left( \xi_D^2\left[ \frac{\nabla}{i} -{\bf A} \right]^2 
+U({\bf r})  \right) \Psi \,,~~~~
\label{WaveEquation}
  \\
U({\bf r}) =-  \xi_D^2(\nabla\alpha)^2  - \sin^2\alpha \sin^2 \mu~~,~~{\bf A} 
({\bf r})=\nabla\alpha\,,~~~~
\label{Potentials}
\\
  \omega_{\rm sat} - \omega_{\rm L} =  \frac{\Omega_P^2}{2\omega_L} (-\lambda \sin^2\mu + \cos^2 \mu)  \,. ~~~~
\label{FrequencyShift}
\end{eqnarray}

For a pure infinite soliton in the $(y,z)$-plane one has $\alpha(x=+\infty)- \alpha(x=-\infty)=\pi$. In this arrangement the vector potential ${\bf A}(x)=\nabla\alpha(x)$ can be gauged away by the transformation $\Psi = e^{i\alpha(x)} \tilde\Psi$. Then one obtains $\lambda=1$, which reflects the mathematical connection of the localised mode to another collective bosonic mode trapped in the soft core of the soliton --- the zero mode corresponding to the translation of the soliton with the eigenfunction $\tilde\Psi(x)= d\alpha/dx$ \cite{vollhardt2013superfluid}.

In the $\xi_D$-vicinity of a half-quantum vortex $\alpha \approx \varphi/2+$ const. ($\varphi$ is the azimuthal coordinate).  Therefore, potential $U(r \leq \xi_D) \propto 1/(4 r^2)$, where $r$ denotes the distance from the HQV core. This would lead to a ground state much more shifted from the bulk line than the one observed, and the actual position would depend on the precise way $U$ is cutoff at $r \sim \xi$ inside the vortex core. This large shift of the order of $\xi_d^2/\xi^2$ is, however, compensated by the so-called Aharonov-Bohm effect, which is manifested by the $\pi$ phase jump of the wave function around the vortex core. The analogy to the original Aharonov-Bohm effect can be seen by introducing an effective magnetic field concentrated in the vortex cores \cite{SalomaaVolovik1987}:
\begin{equation}
{\bf B}_{\rm eff}= \nabla \times {\bf A} =2 \pi \nu \hat{\bf z} 
\delta({\bf r})\,.
\label{ABfield}
\end{equation}
Consequential additional scattering for spin waves leads to a reduced shift of the satellite line in the NMR spectrum. 

\begin{figure}[tb!]
\centerline{\includegraphics[width=0.95 \linewidth]{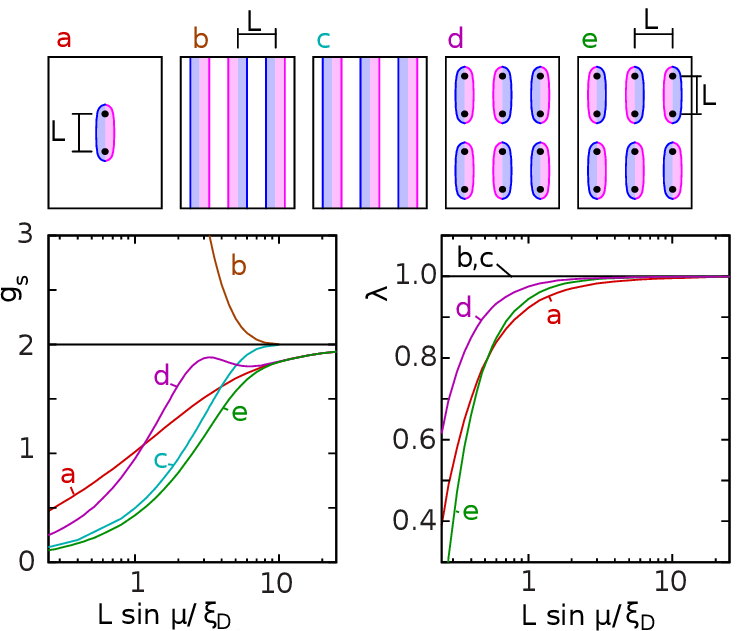}}
\caption{\label{image:numerical} Calculated values of $g_s$ (\emph{bottom left}) and $\lambda$ (\emph{bottom right}) as a function of  $L \sin \mu /\xi_D$. We consider five cases: (a) a single soliton of a finite length L, periodic structures of infinite solitons layers with the period L and same (a) or alternating
soliton orientations (b), and square lattices of solitons and half-quantum vortices with periodic boundaries such that the soliton orientations are alternating (d) and not alternating (e).}
\end{figure}

In general, the satellite peak position depends on the aspect ratio of the soliton $L/D$, where
$D$ is the thickness of the soliton.  In the tilted magnetic field the thickness of the soliton scales as $D \sim \xi_D / \sin\mu$ and thus the finite size effect becomes important at smaller $\mu$. We calculated $\lambda$ numerically as a function of $L\sin\mu/\xi_D$ in a 2D finite element model, i.e., assuming infinite straight vortices and uniform polar phase. Five different cases were studied (Fig.~\ref{image:numerical}). Periodic structures of infinite soliton layers yield $\lambda \equiv 1$, as expected for any unpinned soliton structure. Periodic square-lattice structures of solitons terminating to half-quantum vortices, and a single soliton between two HQVs all yield roughly the same $\lambda(\mu)$. Details of the calculation will be published separately \cite{Slazav_solitons}.

The relative intensity of the satellite peak is obtained using the eigenfunction of the bound state in the soliton (see Eqs. (13) and (39) in Ref.~\citenum{HuMaki1987}). We get
 
 \begin{equation}
I_{\rm sat} =\frac{1}{2}  n_v \frac{\left|\int \Psi 
d^2r\right|^2}{\int |\Psi|^2 d^2r}= \frac{\xi_D^2}{2 \sin^2 \mu} n_v I^{\rm M} . 
\label{Intensity}
\end{equation}
Here $I^{\rm M}=g_s L \sin \mu / \xi_D$ where the numerical factor $g_s$ depends on, e.g., the shape of the wave function. The equilibrium density of vortices is

\begin{equation}
n_v=\frac{ 2m_3\Omega}{\nu \pi \hbar}\,.
\label{VortexDensity}
\end{equation}
The distance $L$ between the HQVs is given by $L^2= b^2/n_v$, where $b=1$ for a square lattice used in the simulations. In a transverse field the relative satellite intensity from the HQVs becomes

\begin{equation}
I_{\rm{sat}} = \frac{g_s}{2} \xi_D  \sqrt{n_v}.
\end{equation}
For $\xi_D=17~\mu$m our numerical simulation yields $g_s \approx 1.7$ independent of the configuration.

\section*{Topology and Superfluid phases of $^3$He}

Many exotic topological objects have been observed experimentally in the superfluid phases of $^3$He, among them quantized vortices with spontaneous asymmetry in the vortex core,  skyrmions with different topological numbers, soliton terminating on the combined spin-mass vortices, and vortex sheets of different geometry. Some exotic objects are still to be detected, such as a nexus -- a Dirac monopole terminating a vortex line, or a boojum -- topological defect living on the surface of the container or at the interface between different superfluid phases \cite{VolovikBook}. 

The known phases of liquid $^3$He belong to 4 different topological classes:

({\bf i}) The normal liquid $^3$He belongs to the class of systems with topologically protected Fermi surfaces. The Fermi surface is described by the first odd Chern number in terms of the Green's function \cite{VolovikBook}.

({\bf ii}) Superfluid $^3$He-A and $^3$He-A$_1$ are chiral superfluids with topologically protected Majorana-Weyl fermions in the bulk. In the relativistic quantum field theories, Weyl fermions give rise to the effect of chiral (axial) anomaly. The direct analogue of this effect has been experimentally demonstrated in $^3$He-A \cite{Bevan1997}. It is the first condensed matter system where the chiral anomaly effect has been observed.  The singly quantized vortices in superfluids with Weyl points contain dispersionless band (flat band) of Andreev-Majorana fermions in their cores  \cite{KopninSalomaa1991}. Discoveries of Weyl fermions have been recently reported  in the topological semiconductors \cite{Huang2015,Lu2015}.

({\bf iii}) $^3$He-B is the purest example of a fully gapped DIII class superfluid with topologically protected
gapless Dirac fermions on the surface \cite{Mizushima2015}.  Some evidence of gapless fermions on the surface of $^3$He-B has been reported \cite{Muarakawa2011,Bunkov2014}.
  
({\bf iv}) Polar phase, the most recently discovered phase, belongs to the class of fermionic materials with topologically protected lines of nodes, and thus two-dimensional flat bands of Andreev-Majorana fermions cover its surfaces \cite{Schnyder2011}. (See also recent reviews  on superconductors with topologically protected nodes \cite{SchnyderBrydon2015}.) Recently the class of these materials has been joined by the topological semimetals \cite{Yu2015,Kim2015,HekkilaVolovik2015}. The half quantum vortex, which we found  in the polar phase, can be represented as a vortex in a single spin component. The spectrum of bound states experiences condensation of zero modes \cite{volovik_fermion_modes}.

It is possible that with properly engineered nanostructured confinement one may reach also new topological phases of liquid $^3$He including

({\bf i}) The planar phase, which is non-chiral, has Dirac nodes in the bulk, and Fermi arc of Andreev-Majorana fermions on the surfaces \cite{Makhlin2014}.

({\bf ii}) The two-dimensional topological states in the ultra-thin film, including inhomogeneous phases of superfluid $^3$He films \cite{Vorontsov2007}. The films with the $^3$He-A and the planar phase order parameters belong to the 2D fully gapped topological materials, which experience the quantum Hall effect and the spin quantum Hall effect in the absence of magnetic field \cite{VolovikYakovenko1989}.

({\bf iii}) $\alpha$-state, which contains 4 left-handed and 4 right-handed Weyl points in the vertices of a cube \cite{VolovikGorkov1985}. It resembles the high energy physics model system with
8 left-handed and 8 right-handed Weyl fermions in the vertices of a four-dimensional cube \cite{Creutz2008}.

\end{document}